\documentclass[preprint,amsmath,amssymb,amssymb,aps,showpacs,pra]{revtex4-1}
\usepackage{fmtcount} 
\usepackage{graphicx}
\usepackage{dcolumn}
\usepackage{bm}
\usepackage{braket}
\usepackage{mathrsfs}
\begin{document}

\title{Phase dependent interference effects on atomic excitation}

\author{Pankaj K. Jha,$^{1,2,*}$ Hebin Li,$^{1,\dagger}$ Vladimir A. Sautenkov,$^{1,3}$ Yuri V. Rostovtsev,$^{4}$ and Marlan O. Scully$^{1,2}$}

\affiliation{$^{1}$Institute for Quantum Science and Engineering and Department of Physics and Astronomy, Texas A\&M University, College Station,Texas 77843, USA\\
$^{2}$Mechanical and Aerospace Engineering and the Princeton Institute for the Science and Technology of Materials, Princeton University, Princeton, NJ 08544, USA\\
$^{3}$P.N.Lebedev Physical Institute of the Russian Academy of Sciences, Moscow 119991, Russia\\
$^{4}$Department of Physics, University of North Texas, Denton, Texas 76203, USA}
\pacs{42.50.-p, 32.80.-t, 37.10.Jk, 42.50.Gy}

\begin{abstract}
\noindent We present an experimental and theoretical study of phase-dependent interference effects in multi-photon excitation under bichromatic radio-frequency (rf) field. Using an intense rf pulse,
we study the interference between the three-photon and one-photon transition between the Zeeman sub-levels of the ground state of $^{87}$Rb that allows us to determine the carrier-envelope phase
of the fields even for long pulses.
\noindent 

\end{abstract}

\maketitle

\section{Introduction}

Modern lasers can produce ultra-short intense pulses
with only a few cycles of carrier oscillation \cite{brabec00rmp}. 
The carrier-envelope phase (CEP) strongly affects many processes
involving few-cycle pulses. In particular, 
CEP effects on high-harmonic generation
\cite{boham98prl}, strong-field photoionization
\cite{paulus01nature}, the ionization of Rydberg atoms
\cite{gurtler04prl}, the dissociation of HD$^+$ and H$_2^+$
\cite{rudnev04prl2}, and the injected photocurrent in semiconductors
\cite{fortier04prl}, 
have been demonstrated by few-cycle pulses. A stabilized and
adjustable CEP is important for applications such as optical
frequency combs \cite{diddams2000} and 
quantum control \cite{yin1992,shapiro2000,hache1997}. 
The frequency dependence of the CEP effect on bound-bound transitions 
has been studied theoretically and experimentally in 
Rb~\cite{hli10prl} and numerically
in Helium interacting with an ultrashort laser pulse 
of a few cycles~\cite{HeCEP}. 

Several techniques have been developed to control the CEP of femtosecond pulses
\cite{jones2000,baluska02nature}. A crucial step in attaining this control 
is measuring the CEP to provide feedback to the laser system. Promising
approaches use, for instance, photoionization
\cite{verhoef06ol,paulus09nature} and quantum interference in
semiconductors \cite{fortier04prl}. While the measurement and control of the CEP is well established 
for few-cycle laser systems, it is entirely absent for high-power, 
many-cycle lasers. 
Recently, a method has been presented for the measurement
of the absolute CEP of a high-power, many-cycle driving pulse,
by measuring the variation of the XUV spectrum 
by applying the interferometric polarization gating technique 
to such pulses~\cite{tzallas10prl}. 

In this paper, we have studied the CEP effects in the population transfer between two bound atomic states interacting with pulses consisting of many cycles in contrast with few-cycle pulses~\cite{hli10prl}. For our experiment, we use intense rf pulses interacting with the magnetic Zeeman sub-levels of rubidium (Rb) atoms. We have found that, for long pulses consisting two frequencies [see Fig.~1], the CEP of the pulse strongly affects that transfer. The physics of the effect is the following. Excitation of upper level [see Fig.~1] is created by two path: 1) three-photon excitation by the field of frequency $\nu_1$ and 2) one-photon excitation by the field of frequency $\nu_2$. The amplitude of excitation for each of paths depends on the phase of the field. As a result, the level of excitation depends on phases of the fields. It is worth noting here that our scheme has no limitation on the duration of pulses.
\begin{figure}[b]
  \includegraphics[width=0.36\textwidth,height=4cm]{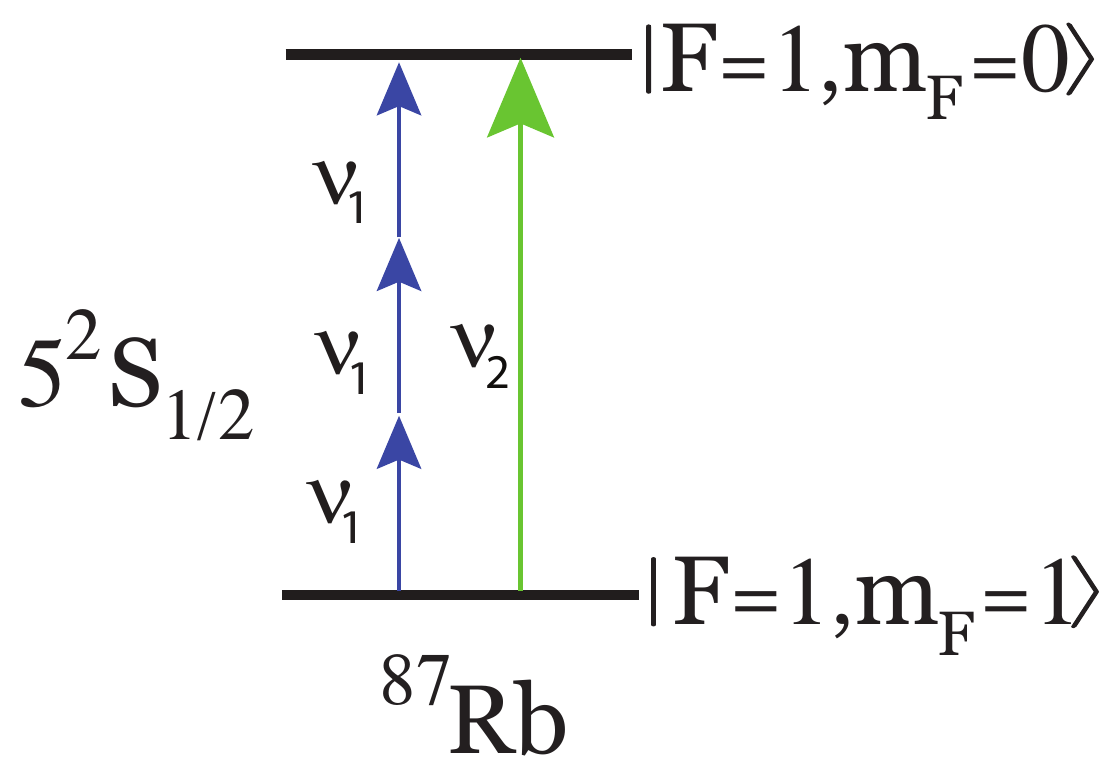}
  \caption{(Color Online) Different channels for excitation between the Zeeman sub-levels of $^{87}$Rb. Atomic transition frequency $\omega=150$kHz.}
\end{figure}
\begin{figure}[b]\label{setup}
  \includegraphics[height=4cm,width=0.35\textwidth,angle=0]{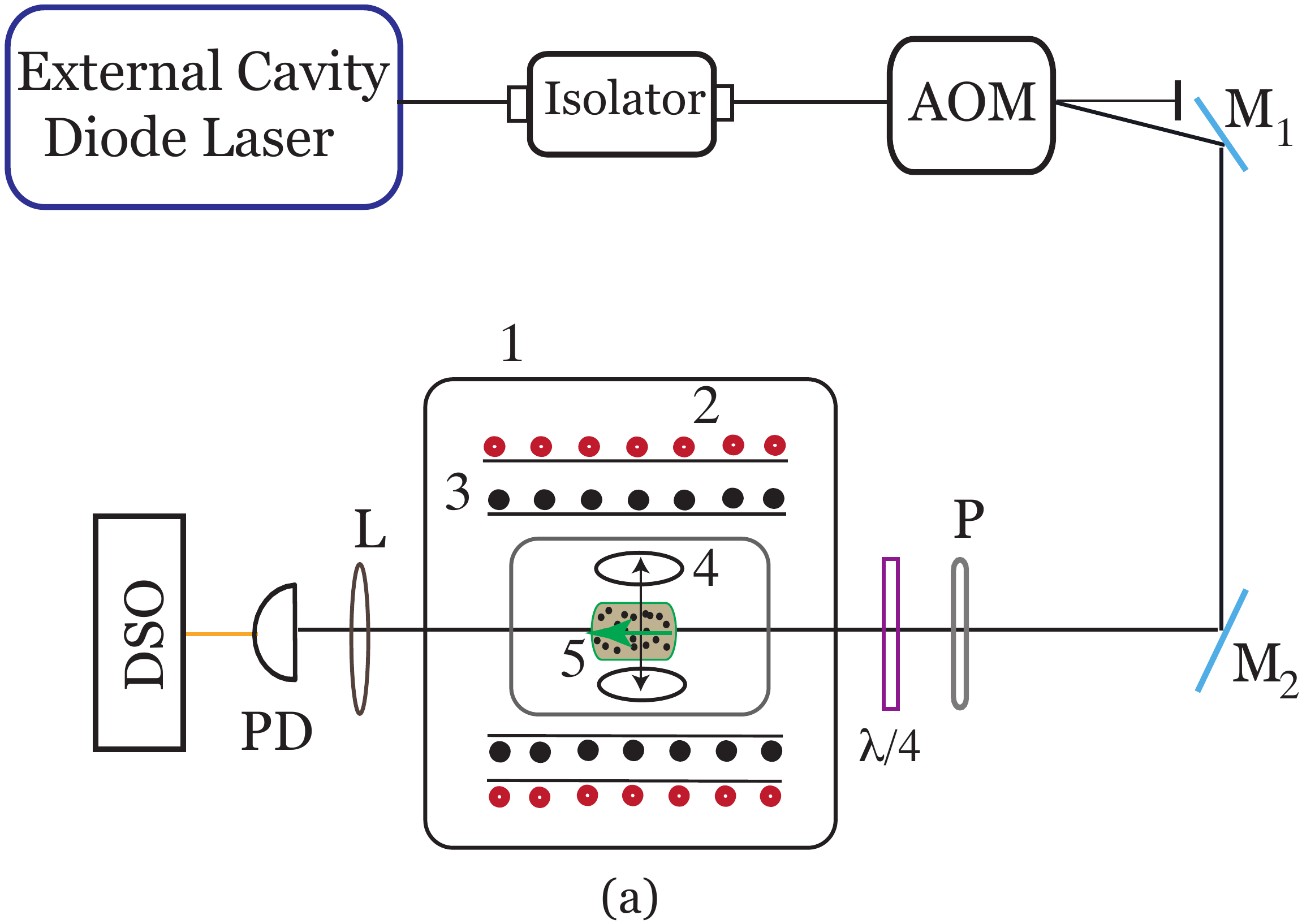}
  \includegraphics[height=4cm,width=0.35\textwidth,angle=0]{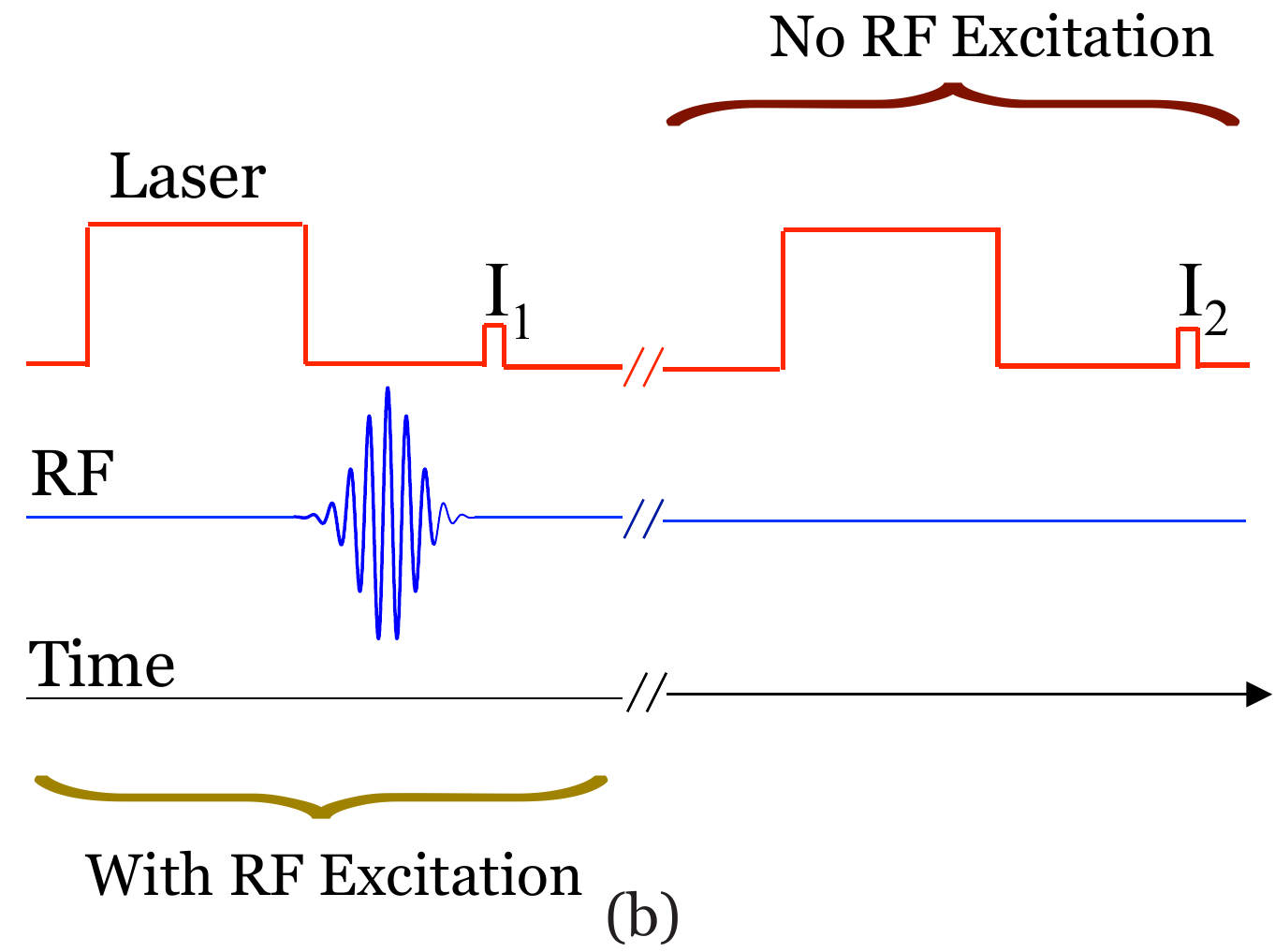}
  \caption{(Color Online) (a) Experimental setup. ECDL-External Cavity Diode Laser; AOM-Acousto-Optic Modulator; P-Polarizer, PD-PhotoDiode; L-Lens, the oven is assembled with 1. Copper tube; 2. Non-magnetic heater on a magnetic shield; 3. Solenoid; 4. Pair of Helmholtz coils; 5. Rb cell. (b) Time sequence of the laser pulses and rf pulse.}
\end{figure}
\section{Experiment}
\subsection{Experimental Setup}
The experimental setup is schematically shown in Fig. 2(a). An external cavity diode laser (ECDL) was tuned to the $\text{D}_{1}$ resonance line of $^{87}$Rb atoms at $| 5^{2}S_{1/2}; F=1\rangle \leftrightarrow | 5^{2}P_{1/2}; F=1\rangle$ transition. A 2.5 cm long cell containing $^{87}$Rb (and 5 torr of Neon) is located in an oven. The cell is heated in order to reach an atomic density of the order of $10^{11}$ cm$^{-3}$. A longitudinal static magnetic field is applied along the laser beam to control the splitting of the Zeeman sub-levels of the ground state $| 5^{2}S_{1/2}; F=1,m_{F}=-1,0,1\rangle$. A pair of Helmholtz coils produces the transverse bichromatic rf magnetic field  with central frequencies at $\nu_{1}=$50 kHz and $\nu_{2}=$150 kHz~\cite{PS}. In our experiment the function generator was programmed to provide multi-cycle bichromatic pulses with controllable parameters, such as the pulse duration, CEPs and the amplitudes of the two carrier frequencies.

The energy level scheme for $^{87}$Rb is shown in Fig. 3(b). The ground state has three sub-levels; a right-circularly polarized laser pulse optically pumps the system and drives the atoms to the sub-level $| 5^{2}S_{1/2}; F=1,m_{F}=1\rangle$. This is followed by the bichromatic rf pulse, which excites the atoms to the sub-levels $| 5^{2}S_{1/2}; F=1,m_{F}=-1,0\rangle$. The population in these sub-levels is subsequently determined by measuring the transmission of a weak right-circularly polarized probe pulse. 
\begin{figure*}[t]
  \includegraphics[width=0.80\textwidth,height=5.5cm]{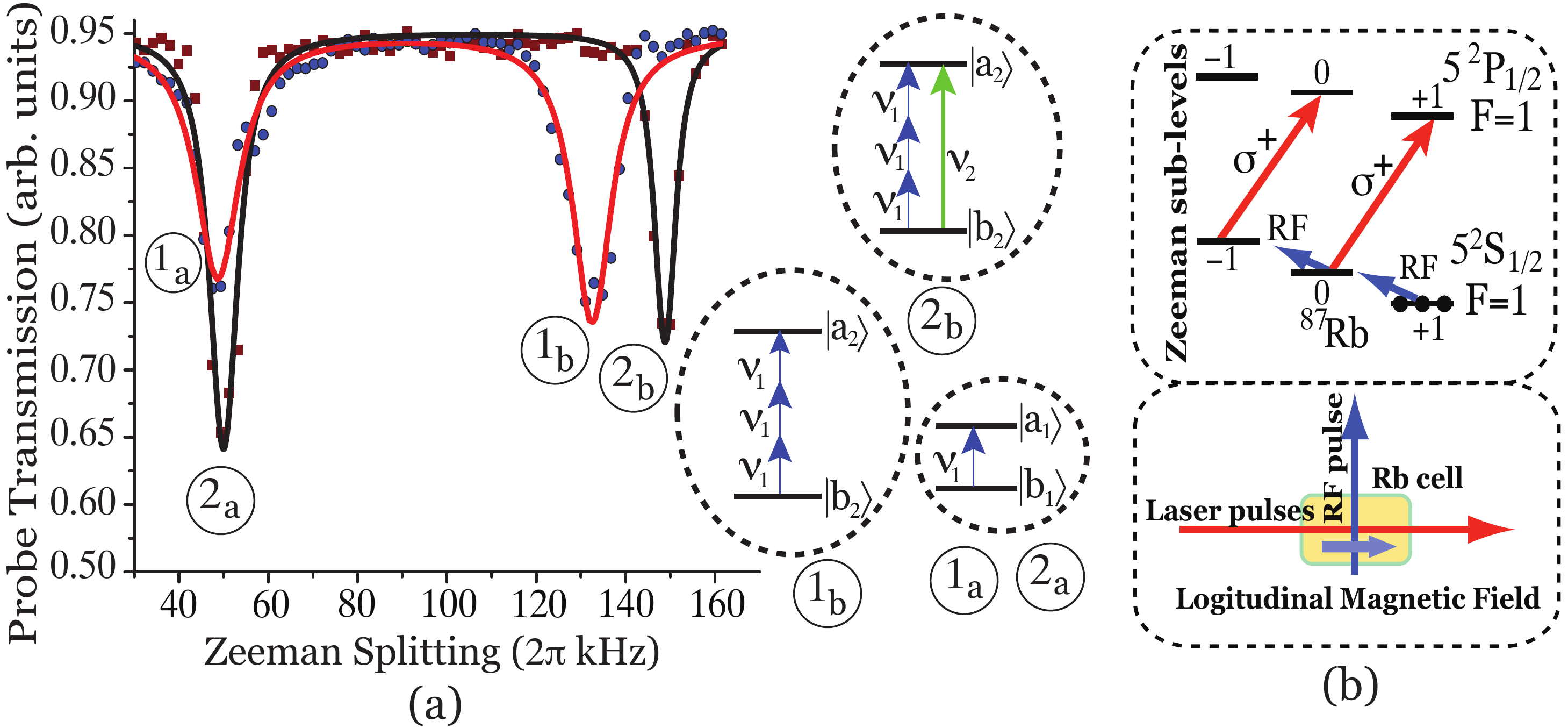}
  \caption{(Color Online) (a)Transmission profile of the weak optical probe field. Peak $1_{a}$ and $2_{a}$ corresponds to one-photon absorption while peak $1_{b}$ and $2_{b}$ correspond to three-photon absorption. Peaks $2_{a,b}$ corresponds to bichromatic rf field excitation while peaks $1_{a,b}$ corresponds to the excitation when the amplitude for 150kHz components is zero. The dots (solid circles) and square boxes are the experimental results while the solid curves are the best fit. (b) Energy level (upper block) and the geometry of the Rb cell and the applied fields (lower block). }
\end{figure*}
To determine the population transfer due to the rf excitation, the experiment is performed with a sequence of laser pulses with a rf pulse followed by a sequence of laser pulses without rf pulse. For the transmitted probe pulse, intensity is given by $I_{1}=I_{0}\eta e^{N\sigma LP_{a}}$, where $I_{0}$ is the probe pulse input intensity, $\eta$ is the factor due to dephasing, $N$ is the atomic density, $\sigma$ is the absorption cross-section, $L$ is the cell length and $P_{a}$ is the population of the upper levels due to rf excitation. For the second sequence , in which there is no rf excitation, the transmitted probe pulse intensity is given by $I_{2}=I_{0}\eta$. Therefore, the population due to rf excitation is given by the quantity $-ln(I_{1}/I_{2})=N\sigma L P_{a}$. 

The time sequence of the laser pulses and the rf pulse is shown in Fig. 2(b). The sequence of the laser pulse includes a 1.0 ms strong pulse to optically pump the atoms and a 2$\mu s$ weak pulse (delayed 330 $\mu s$) to probe the populations of the upper Zeeman sub-levels. The bichromatic rf pulse is delayed by 165$\mu s$ with respect to the optical-pumping laser and its duration, T (full width at half maximum, FWHM), is 130 $\mu s$. The transmitted intensity of the probe pulse is monitored by a fast photodiode.
\subsection{Experimental Results}
In Fig. 3(a) we have shown the single-photon and multi-photon excitation under bichromatic field interaction with $^{87}$Rb. Peak $(1_{b})$ is single-photon absorption peak at frequency $\omega_{1}=$50kHz. Peak $(2_{b})$ corresponds to the three-photon absorption of $\nu_{1}$=50kHz and single-photon absorption of $\nu_{2}=$150kHz. Peak $(2_{b})$ emerges due of two different paths of excitation between the initial and the final states [inset Fig. 3(a)]. On the other hand when the amplitude for 150 kHz component is reduced to zero, the three-photon absorption peak shifts to the left around $130$ kHz [see peak $1_{b}$]. This kind of shift was studied by Ramsey \cite{J3} in mid 50's and he showed that  if resonance transitions are induced by a perturbation at one frequency, then the presence of the other perturbations at non-resonant frequencies alters the resonance frequency of the first perturbations. Bloch-Siegert shift~\cite{BS} is a special case of this general phenomena. We could see that in case of the two central frequency pulse, the shift is less and peak is close to $\omega=150$kHz. In this paper we investigated the phase dependent peak $2_{b}$ by varying the phase of $\phi_{150}$ while keeping the phase $\phi_{50}=0$. In our experiment we controlled the Zeeman splitting by tuning the longitudinal magnetic field applied along the laser (as shown in Fig 3(b) lower block) and keeping the frequencies of the rf pulse intact.

The main result of the experiment is shown in Fig. 4. In Fig. 4, we have shown the transmission profile of the weak optical probe field (which measure the populations of the upper Zeeman sub-levels) as function of the Zeeman-splitting. Interestingly we see a  single peak $2_{\alpha}$ for $\phi_{150}=\phi_{50}=0$ and two well defined peaks $2_{\alpha}$ and $2_{\beta}$ when the phase $\phi_{150}=2\pi/3$. Existence of the second peak can be clearly observed around $\phi_{150}=\pi/3$. When the two frequency components are completely out of phase, the first peaks disappears and we have the single peak $2_{\beta}$. Thus in the interval $0 \le \phi_{150} \le \pi$, the profile of probe transmission changes from peak $2_{\alpha}$ to $2_{\beta}$ [see Fig 5(a)].
\begin{figure*}[htb]
 \includegraphics[height=9cm,width=0.98\textwidth]{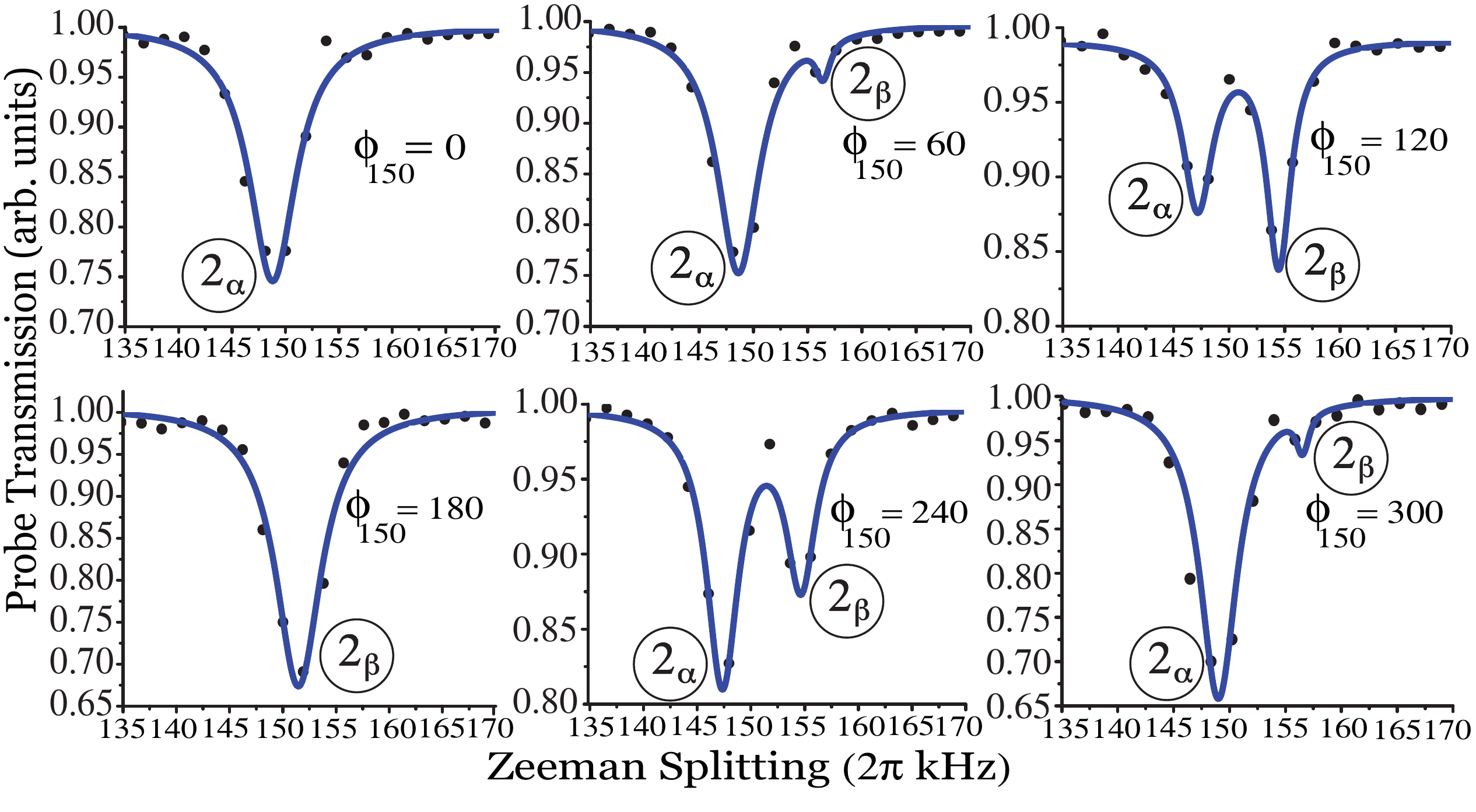}
    \caption{(Color online) Transmission profile of the weak optical probe field under bichromatic rf field excitation for different values of the absolute phase of $\phi_{150} =0 \rightarrow 360$ keeping $\phi_{50}=0$. The dots (solid circles) are the experimental results while the solid curves are the best fit. Unit of phase is degree here.  }
\end{figure*}
\section{Theory}
The Hamiltonian for an atomic state with F=1 in a magnetic field $B=(B_{x},B_{y},B_{z})$ is given by
\begin{equation}\label{e1}
\mathscr{H}=-g\mu_{0}\left( {\begin{array}{ccc}
 B_{z} & \frac{B_{x}+iB_{y}}{\sqrt{2}} & 0  \\
 \frac{B_{x}-iB_{y}}{\sqrt{2}} & 0 & \frac{B_{x}+iB_{y}}{\sqrt{2}}   \\
 0 & \frac{B_{x}-iB_{y}}{\sqrt{2}} & -B_{z}  \\
 \end{array} } \right),
\end{equation}
where $g=-1/2$ is the Lande factor for this Rb state, $\mu_{0}$ is the Bohr magneton, the $B_{z}=B_{0}$ is the static magnetic field that is chosen in the direction of the z-axis; the $B_{x}$ and $B_{y}$ are the transverse components driven by a function generator. The linearly-polarized bichromatic magnetic field is given as,
\begin{equation}\label{e2}
B_{x}(t)=e^{-\alpha ^{2}t^{2}}\{B_{1}\text{cos}(\nu_{1}t+\phi_{1})+B_{2}\text{cos}(\nu_{2}t+\phi_{2})\},
\end{equation}
where $\alpha = (2\sqrt{\text{ln}2})/T$ and T is the FWHM duration of the pulse and $B_{y}=0$.
For the magnetic dipole transition, the relaxation due to atomic motion is the most important. The density matrix equations is given by
\begin{equation}\label{e3}
\dot{\rho}=-\frac{i}{\hbar}[\mathscr{H},\rho]-\Gamma (\rho-\rho_{0}),
\end{equation}
where $\mathscr{H}$ is given by Eq.(\ref{e1}), $\Gamma$ quantifies the relaxation process due to atomic motion and $\rho_{0}$ is the thermal equilibrium density matrix of the atoms in the cell without the optical and RF fields. For simple explanation we will consider only two levels coupled by the bichromatic field and neglect any type of relaxation. The Rabi frequency is given by
\begin{equation}\label{e4}
\Omega(t)=2e^{-\alpha ^{2}t^{2}}\{\Omega_{1}\text{cos}(\nu_{1}t+\phi_{1})+\Omega_{2}\text{cos}(\nu_{2}t+\phi_{2})\},
\end{equation}
where $\Omega_{(1,2)}=g\mu_{0}B_{(1,2)}/2\sqrt{2}\hbar$. The equation of motions for the probability amplitudes $C_{a}$ and $C_{b}$ are given by,
\begin{subequations}\label{e5}
\begin{align}
\dot{C}_{a}&=i\Omega(t)e^{i\omega t}C_{b}, \label{second}\\
\dot{C}_{b}&=i\Omega^{*}(t)e^{-i\omega t}C_{a}.
\end{align}
\end{subequations}
Let us consider the perturbative approach $C_{b}(t) \cong 1$. We look for a solution of the form $C_{a}(\infty)=C_{a}^{(1)}(\infty)+C_{a}^{(3)}(\infty)$, where the first term is given by
\begin{equation}
C_{a}^{(1)}(\infty)=i\int_{-\infty}^{\infty}\Omega(t')e^{i\omega t'}dt'.
\end{equation}
\begin{figure*}[htb]
  \includegraphics[width=0.75\textwidth,height=5.5cm]{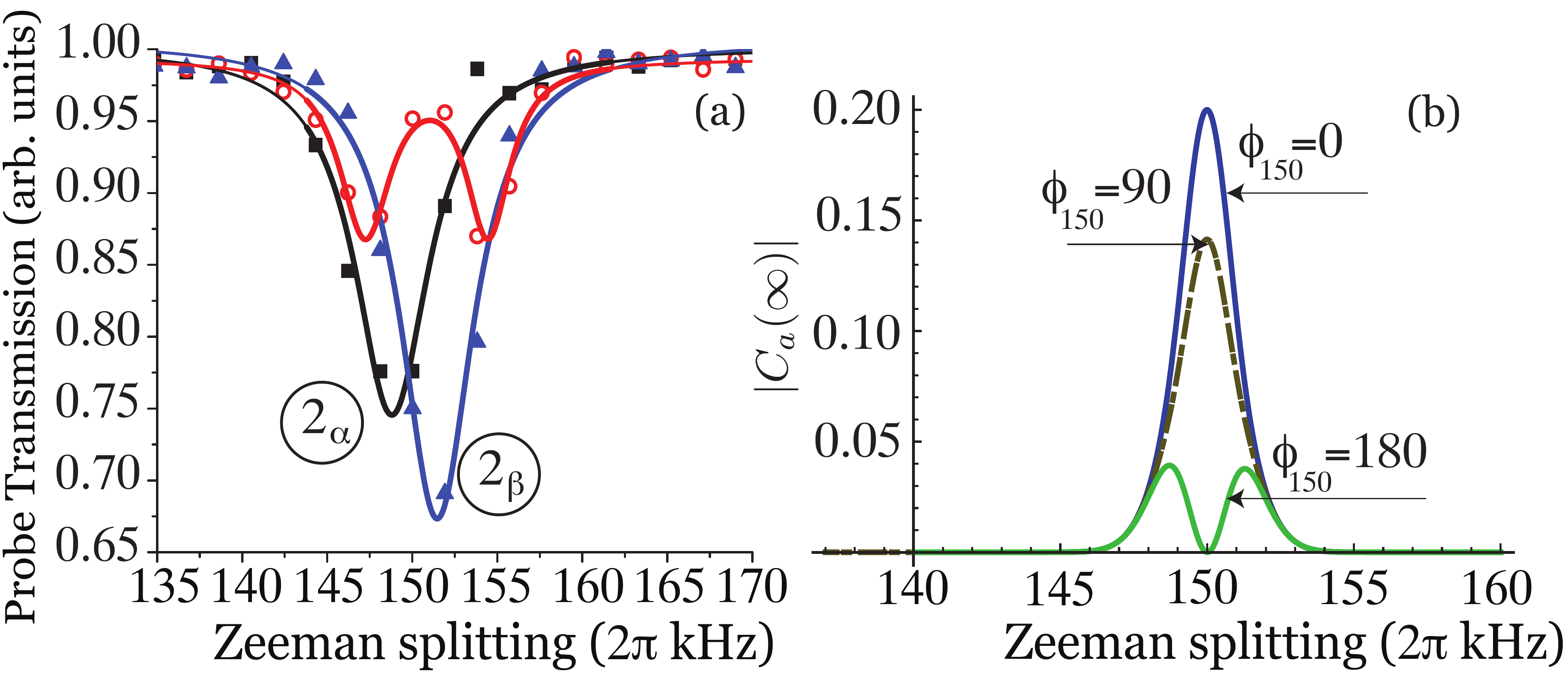}
  \caption{(Color Online) Transmission profile of the weak optical probe field for different values of the absolute phase of $\phi_{150} $ keeping $\phi_{50}=0$. $\phi_{150} =0$ (solid square Black line), $\phi_{150}=115$ (hollow circle Red line) and $\phi_{150} =180$ (solid triangle Blue line). (b) $|C_{a}(\infty)|$ for three combination of $\phi_{150} $ for $\phi_{50}=0$ using Eq.(\ref{e14}). Unit of phase is degree here.}
  \label{fig6}
\end{figure*}
The second term can be found as
\begin{equation}
\begin{split}
C_{a}^{(3)}(\infty)=-i\int_{-\infty}^{\infty}\left\{\Omega(t)e^{i\omega t}\int_{-\infty}^{t}\left[\Omega^{*}(t'')e^{-i\omega t''}
\int_{-\infty}^{t''}\Omega(t')e^{i\omega t'}dt'\right]dt''\right\}dt.
\end{split}
\end{equation}
Here $C_{a}^{(1)}(\infty)$ and $C_{a}^{(3)}(\infty)$ quantifies the probability of one-photon and three-photon absorption. Using simple algebra we can find $C^{(1)}_{a}(\infty)$
\begin{equation}\label{e9}
C_{a}^{(1)}(\infty)=i\left(\frac{\sqrt{\pi}}{\alpha\hspace{0.5mm}
}\right)\Omega_{2}\hspace{1mm}e^{-\left[(\omega -\nu_{2})/2\alpha\right]^{2}}e^{-i\phi_{2}},
\end{equation}
and the second term can be found as
\begin{equation}\label{e13}
C_{a}^{(3)}(\infty)=-i\left\{\frac{\sqrt{\pi}}{2\sqrt{3}\hspace{0.5mm}\alpha\hspace{0.5mm}\nu_{1}(\omega-\nu_{1})
}\right\}\Omega^{3}_{1}\hspace{1mm}e^{-\left[(\omega -3\nu_{1})^{2}/12\alpha^{2}\right]}e^{-3i\phi_{1}}.
\end{equation}
Combining Eq.(\ref{e13}) and Eq.(\ref{e9}) we obtain 
\begin{equation}\label{e14}
\begin{split}
C_{a}(\infty)=i\left(\frac{\sqrt{\pi}}{\alpha\hspace{0.5mm}
}\right)\Omega_{2}\hspace{1mm}e^{-\left[(\omega -\nu_{2})/2\alpha\right]^{2}}e^{-i\phi_{2}}
-i\left\{\frac{\sqrt{\pi}}{2\sqrt{3}\hspace{0.5mm}\alpha\hspace{0.5mm}\nu_{1}(\omega-\nu_{1})
}\right\}\Omega^{3}_{1}\hspace{1mm}e^{-\left[(\omega -3\nu_{1})^{2}/12\alpha^{2}\right]}e^{-3i\phi_{1}}.
\end{split}
\end{equation}
The amplitude $|C_a(\infty)|$ depends on the phases $\phi_1$ and $\phi_2$ as is clearly seen from Eq.(\ref{e14}), and as is shown in Fig. 5(b). Let us underline here that the range of frequencies where this interference occurred depends on the parameter $\alpha$ related to the pulse duration, but the effect of interference does not depend on $\alpha$. Thus our scheme has no limitation on the duration of pulses. 

\section{Conclusion}

In conclusion, for our experiment, we use intense rf pulses interacting with the magnetic Zeeman sub-levels of rubidium (Rb) atoms, we have experimentally and theoretically shown the CEP effects in the population transfer between two bound atomic states interacting with pulses consisting of many cycles (up to 15 cycles) of the field. We have found that, for long pulses consisting two carrier frequencies, the CEP of the pulses strongly affects that transfer. The significance of our experiment is that it provides the insight of CEP effect in a new regime. Our study in the RF domain suggests new experiments with bound states and laser pulses in the optical domain that furnishes another way to measure the CEP. Furthermore, the observed phase dependent excitation, as a result of the interference between one- and three-photon transitions \cite{fortier04prl,peng2008}, is important to quantum-control experiments \cite{yin1992,shapiro2000,hache1997}. Our experiment provides a unique system serving as an experimental model for studying ultrashort optical pulses.The obtained results may be easily extended to optical experiments.

\section{Acknowledgement}
We thank  Leonid V. Keldysh, Olga Kocharovskaya, M.S.Zubairy and T. Siebert  for useful discussions and gratefully acknowledge the support from the NSF Grant EEC-0540832 (MIRTHE ERC), Office of Naval Research (N00014-09-1-0888 and N00014-08-1-0948), Robert A. Welch Foundation (Award A-1261)) and the partial support from the CRDF. P.K.J would also like to acknowledge the Robert A. Welch Foundation and HEEP Foundation for financial support, and Y.V.R. gratefully acknowledge the support from the UNT Research Initiation Grant and the summer fellowship UNT program.
 
\noindent$^*$Email: pkjha@physics.tamu.edu\\
$\dagger$Current Address: JILA, University of Colorado.


\begin{thebibliography}{99}
\bibitem{PS} Here we use the convention that all frequencies are circular frequencies so that $\hbar\nu$ (not $h\nu$) is the photon energy and the  atomic transition frequency is $\omega=\omega_{a}-\omega_{b}$. We define detuning as $\Delta=\omega-\nu$.
\end{thebibliography}
\end{document}